\documentclass[12pt]{iopart}
\usepackage{iopams}  
\usepackage{graphicx}
\usepackage[normalem]{ulem}
\usepackage[squaren]{SIunits}
\usepackage{xcolor}
\usepackage[breaklinks=true,colorlinks=true,linkcolor=blue,urlcolor=blue,citecolor=blue]{hyperref}
\usepackage{lineno}

\begin{document}

\title[E300 first results]{Generation of meter-scale hydrogen plasmas and efficient, pump-depletion-limited wakefield excitation using 10 GeV electron bunches}

\author{C. Zhang$^{1}$, D. Storey$^2$, P. San Miguel Claveria$^3$, Z. Nie$^1$, K. A. Marsh$^1$, M. Hogan$^2$, W. B. Mori$^{1,4}$, E. Adli$^5$, W. An$^6$, R. Ariniello$^2$, G. J. Cao$^5$, C. Clarke$^2$, S. Corde$^3$, T. Dalichaouch$^4$, C. E. Doss$^7$, C. Emma$^2$, H. Ekerfelt$^2$, E. Gerstmayr$^{2,8}$, S. Gessner$^2$, C. Hansel$^7$, A. Knetsch$^{2,3}$, V. Lee$^7$, F. Li$^4$, M. Litos$^7$, B. O'Shea$^2$, G. White$^2$, G. Yocky$^2$, V. Zakharova$^3$, and Chan Joshi$^{1}$}
\address{$^1$Department of Electrical and Computer Engineering, University of California Los Angeles, Los Angeles, CA 90095, USA}
\address{$^2$SLAC National Accelerator Laboratory, Menlo Park, CA 94025, USA}
\address{$^3$LOA, ENSTA Paris, CNRS, Ecole Polytechnique, Institut Polytechnique de Paris, 91762 Palaiseau, France}
\address{$^4$Department of Physics and Astronomy, University of California Los Angeles, Los Angeles, CA 90095, USA}
\address{$^5$University of Oslo, 0316 Oslo, Norway}
\address{$^6$Department of Astronomy, Beijing Normal University, Beijing 100875, China}
\address{$^7$Department of Physics, Center for Integrated Plasma Studies, University of Colorado Boulder, Boulder, CO 80309, USA}
\address{$^8$Stanford Pulse Institute, Menlo Park, CA 94305, USA}

\ead{chaojiez@ucla.edu}
\vspace{10pt}

\begin{abstract}
High repetition rates and efficient energy transfer to the accelerating beam are important for a future linear collider based on the beam-driven plasma wakefield acceleration scheme (PWFA-LC). This paper reports the first results from the Plasma Wakefield Acceleration Collaboration (E300) that are beginning to address both of these issues using the recently commissioned FACET-II facility at SLAC National Accelerator Laboratory. We have generated meter-scale hydrogen plasmas using time-structured 10 GeV electron bunches from FACET-II, which hold the promise of dramatically increasing the repetition rate of PWFA by rapidly replenishing the gas between each shot compared to the hitherto used lithium plasmas that operate at 1-10 Hz. Furthermore, we have excited wakes in such plasmas that are suitable for high gradient particle acceleration with high drive-bunch to wake energy transfer efficiency- a first step in achieving a high overall energy transfer efficiency. We have done this by using time-structured electron drive bunches that typically have one or more ultra-high current ($>$30 kA) femtosecond spike(s) superimposed on a longer ($\sim$0.4 ps) lower current ($<$10 kA) bunch structure. The first spike effectively field-ionizes the gas and produces a meter-scale (30-160 cm) plasma, whereas the subsequent beam charge creates a wake. The length and amplitude of the wake depends on the longitudinal current profile of the bunch and plasma density. We find that the onset of pump depletion, when some of the drive beam electrons are nearly fully depleted of their energy, occurs for hydrogen pressure $\geq$1.5 Torr. We also show that some electrons in the rear of the bunch can gain several GeV energies from the wake. These results are reproduced by particle-in-cell simulations using the QPAD code. At a pressure of $\sim$2 Torr, simulations results and experimental data show that the beam transfers about 60\% of its energy to the wake.
\end{abstract}

%
%
%
%
%

\section{Introduction}
Plasma wakefield acceleration has emerged as a promising frontier in advanced acceleration research \cite{joshi_new_2021}, offering potential for significantly reducing the size and cost of a future accelerator operating at the energy frontier of high-energy particle physics. In recent years, significant progress has been made in experimental and theoretical investigations of PWFAs. For instance, high-gradient \cite{hogan_multi-gev_2005,blumenfeld_energy_2007}, high-efficiency and narrow energy spread acceleration of a distinct trailing electron bunch \cite{litos_high-efficiency_2014,litos_9_2016}, and narrow energy spread acceleration of positrons \cite{corde_multi-gigaelectronvolt_2015,doche_acceleration_2017} have been demonstrated using the FACET advanced acceleration test facility at the SLAC National Accelerator Laboratory (SLAC). A similar program is underway at the FLASHForward facility at DESY using a high brightness 1-GeV electron bunch, making progress in high-resolution wakefield measurement \cite{schroder_high-resolution_2020-1} and energy spread preservation in PWFA \cite{lindstrom_energy-spread_2021}.

Most past PWFA experiments at SLAC have used either a laser ionized or a beam ionized lithium plasma source \cite{oconnell_dynamic_2002,muggli_photo-ionized_1999,muggli_meter-scale_2004,blue_plasma-wakefield_2003,gessner_demonstration_2016}. Beam-ionized argon plasmas have also been used in the earlier FACET experiments \cite{corde_high-field_2016}. The wakes are produced by either 4 ps (1.2 mm) long $e^-$ or $e^+$ bunches (laser produced plasma) or highly compressed 15-50 $\rm\mu m$ $e^-$ or $e^+$ drive bunches (beam-ionized plasmas). The low ionization potential of Li (5.4 eV) allows for the formation of a partially ionized low density ($<2\times10^{14}~\rm{cm^{-3}}$) plasma using laser ionization or a high density ($<2\times10^{17}~\rm{cm^{-3}}$) fully beam-ionized meter-scale plasma columns for peak drive-bunch (henceforth referred to as the drive beam, driver or beam) current of $>8$ kA (for $\sigma_r=30~\rm\mu m$).  In the laser ionization case, the beam creating a wake has to be aligned to propagate on the axis of the plasma column whereas in the beam-field ionization case the plasma axis and the beam axis are self-aligned. While this source has proved to be remarkably reproducible and robust in a number of groundbreaking PWFA experiments (in addition to those mentioned above, see references \cite{wang_x-ray_2002,oz_ionization-induced_2007,johnson_positron_2006,gessner_demonstration_2016,lindstrom_measurement_2018,corde_high-field_2016,clayton_self-mapping_2016-1,doche_acceleration_2017}), it is found to be a limiting factor in high repetition rate experiments beyond 1 Hz in continuous wave (CW) mode or 10 Hz in a burst mode operation due to accumulative heating of the plasma. Furthermore, since the lithium vapor column is confined in a high temperature heat pipe oven \cite{muggli_photo-ionized_1999} enclosure by buffer helium gas, there is no access to the plasma in the transverse direction for diagnostic purposes. These difficulties could potentially be overcome if a hydrogen plasma could be used instead of lithium plasma. The main difficulty is that the higher ionization potential (IP) of the hydrogen molecule (15.4 eV) requires either a high intensity ($>5\times10^{14}~\rm{W/cm^2}$), multi-TW laser beam \cite{zhang_ionization_2021} or an electron beam with a peak current $>30$ kA for beam-induced ionization. We note that at the present time we have not considered even lower IP noble gases such as Ar, Kr, Xe because of possible beam-induced multiple ionization that is known to inject a substantial dark current into the wake \cite{vafaei-najafabadi_beam_2014,vafaei-najafabadi_evidence_2016} thereby increasing the complexity of a future multi-stage collider.

Even if one could trigger hydrogen ionization locally, the next challenge is how to produce a long enough plasma to support a meter-scale wake such that most of the drive bunch energy can be coupled to the wake for achieving a high overall efficiency. Currently, work is in progress to produce meter-scale columns of hydrogen plasma using laser ionization \cite{ariniello_laser_2018} and discharge in a sapphire capillary \cite{pena_energy_2023}. One key experiment undertaken at the earlier FFTB facility at SLAC showed that an ultra-high (50 GeV/m) gradient wake can be sustained by a self-guided beam in a meter-scale lithium plasma \cite{blumenfeld_energy_2007}. In this experiment the wake excited by a 42 GeV bunch was ultimately terminated by a process called ``head erosion'' long before the drive bunch was substantially depleted of its energy. Head erosion refers to the phenomenon where the position of the ionization front in a beam ionized plasma recedes backwards until the beam can no longer produce a wake \cite{an_strategies_2013}. Here we get around the head-erosion limited beam propagation distance by reducing the drive bunch energy to 10 GeV and using a relatively gentle beam focusing optics ($\beta^*\geq50~\rm cm$) such that the head of the bunch ahead of the ionization front diverges slowly, causing the ionization front to erode backwards slowly, thereby increasing the pump depletion length \cite{corde_high-field_2016}. Such beams and focusing optics are now available at FACET-II, a state-of-the-art 10 GeV electron beam facility for advanced accelerator research \cite{yakimenko_facet-ii_2019} that came online in 2022 at the SLAC National Accelerator Laboratory.

The FACET-II facility is designed to deliver high-brightness electron bunches with extreme parameters (e.g., $I_{\rm peak}>50$ kA) for advanced beam physics research. It has been experimentally demonstrated that at the present stage of commissioning of the FACET-II accelerator, the production of highly compressed, low emittance beams can result in large shot-to-shot variation in the current profile after the final stage of bunch compression due to radio-frequency (RF) jitters and microbunching instability \cite{ratner_time-resolved_2015}. This leads to a varying time-structured (not the usual single-Gaussian) pulse that features one or multiple ultra-short ($<10$ fs) high peak current spikes. When focused by the final focusing quadrupoles prior interaction point, the very first spike can generate intense transverse electric fields capable of ionizing higher IP molecular gases such as hydrogen over meter-scale lengths, whereas a second much longer ($\sim$0.4 ps) but lower current bunch structure that follows the ionization front excites a wake in the plasma. In a recent paper \cite{storey_wakeeld_2023} we have described the various diagnostics that we employ to characterize the longitudinal phase space of the compressed electron bunch. These diagnostics have shown that the 10 GeV electrons delivered by the linac, operating in the single-bunch mode, have one or more current spikes but neither their amplitude nor frequency can be presently resolved. Nevertheless, in the past we have found that plasma formation in gases with different IPs and the formation of non-evolving wakes as inferred from the energy loss of the drive beam as a function of plasma density and length is ultimately the best indicator of the beam brightness.

Another major innovation since the earlier experimental campaigns on the FFTB and FACET is the development of the quasi-static PIC code QPAD \cite{li_quasi-static_2021}. QPAD is particularly suitable for rapidly carrying out a large number of simulations of the experimental outcomes as it effectively combines the quasi-static approximation and azimuthal decomposition of the electromagnetic fields, considerably accelerating the simulation process and making it feasible to model meter-scale PWFA. This code in practice is typically an order of magnitude more efficient than the previously used QuickPIC \cite{an_improved_2013} and 2-3 orders of magnitude more efficient than the fully self-consistent OSIRIS \cite{goos_osiris_2002}. At present the SLAC linear accelerator infrastructure allows experiments to be carried out at a relatively high repetition rate (up to 30 Hz), while rapid turnaround of the QPAD simulations makes it possible to simulate a large number of experimental outcomes using the measured beam parameters for a particular group of shots. In this experiment we have used QPAD extensively for the first time to generate possible outcomes of the experiment using the beamline simulations. This has allowed us to identify the conditions that are needed to, i) generate meter-scale plasmas in hydrogen, ii) pump deplete the 10 GeV drive bunch, and iii) determine the energy transfer efficiency of the portion of the drive bunch charge that is available to form the wake.

In this paper, we show that it is possible to generate a high-gradient wake in meter-scale beam-ionized hydrogen plasma until some of the 10 GeV drive bunch electrons are nearly fully depleted of their energy while transferring their energy to plasma wake. Demonstration of pump depletion is one of the major goals of the E300 experiment \cite{joshi_plasma_2018}. This onset of pump depletion occurs at a hydrogen gas pressure of 1.5 Torr. At a higher pressure of 2.2 Torr more electrons approach full pump depletion and nearly every shot shows pump depletion. Despite the large shot-to-shot differences due to the variation in the peak current and the current profile of the drive bunch, our analysis and PIC simulations reveal that the energy transfer efficiency of the portion of the drive bunch that follows the high current ionizing spike to the wake reaches about 60\%. This efficiency is observed for a plasma density of $\sim7\times10^{16}~\rm{cm^{-3}}$ (pressure of $\sim2.2$ Torr) and a plasma length of 50 cm. Furthermore, at this pressure we also observe the energy gain of the tail electrons of several GeV.

\section{Beamline and PIC simulations}
\subsection{Generation of time structured electron drive bunch and plasma formation}
Apart from the standard beam diagnostics found on an accelerator beamline, such as wire scanners, beam position monitors and toroidal charge monitors that give information about the beam size, position, and charge, the FACET-II facility has advanced beam diagnostics such as electro-optic sampling (EOS) \cite{hunt-stone_electro-optic_2021}, and an X-band transverse deflecting cavity (X-TCAV) \cite{krejcik_commissioning_2013} that measure beam parameters like the length of a single bunch and separation between drive and trailing bunches. When the beamline is set up for the high compression mode of beam delivery, these parameters are extremely sensitive to RF amplitude and phase jitter in the linac and can fluctuate from shot to shot. During the current ramp up phase of the FACET-II facility not all the diagnostics were operational for the experimenters. We therefore use comprehensive start-to-end beamline simulations to predict the crucial beam parameters at the interaction point.  Starting from the photocathode injector- modeled using the commercial code General Particle Tracer \cite{sb_van_der_geer_general_2009}- the simulations cover the entire one kilometer long beamline modeled using the code Lucretia \cite{tenenbaum_lucretia_2005}, a Matlab-based particle tracking script encompassing all magnets, RF elements and collimators. Lucretia simulations on the LCLS beamline have shown an excellent agreement with the experimental measurements of the longitudinal phase space of the beam made using the X-TCAV diagnostic \cite{emma_machine_2018}.

\begin{figure*}[htb]
 \centering
 \includegraphics[width=0.7\textwidth]{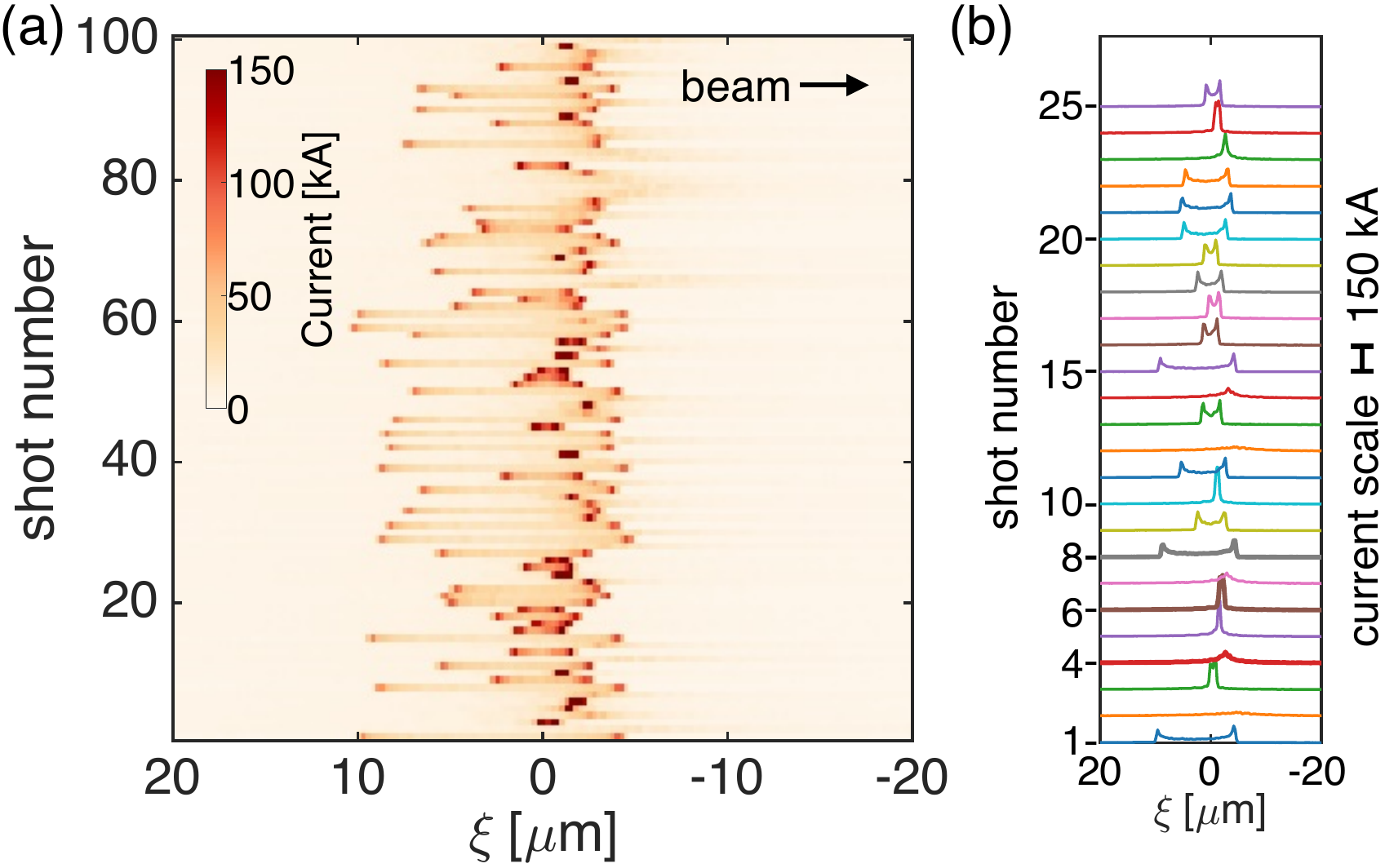}
 \caption{\label{fig1-extreme-beam}Simulated current profile of the drive electron bunches. (a) Beam current profile of 100 beamline simulations showing shot-to-shot fluctuations of the peak current and the position of the current peak when a rms 0.1\% amplitude and 0.25-degree phase jitters of the 2.8 GHz RF are introduced. (b) The longitudinal current profile of the first 25 shots. The lines are vertically shifted to improve clarity.}
\end{figure*}

Among the most crucial beam parameters is the current profile, or the longitudinal phase space profile that directly influences the ionization dynamics of the static-fill hydrogen gas. The optimal compression of the electron bunch after propagation through the final bunch compressor before the final focusing optics is sensitive to the RF amplitude and phase jitter. To investigate this, we introduced in the simulation code a rms 0.1\% jitter of the RF amplitude and 0.25-degree jitter of the RF phase of the main accelerator, and conducted 100 beamline simulations. These numbers are equal to or smaller than the level of RF jitter expected for the SLAC linac. The resulting simulated current profiles at the interaction point are summarized in Fig. \ref{fig1-extreme-beam}(a), with each row representing one independent simulation. To enhance clarity, the current profiles of the first 25 shots are presented in Fig. \ref{fig1-extreme-beam}(b). We can see that there is a great deal of shot-to-shot variation of the current profile even with such a small amount of jitter of the RF amplitude and phase.

\begin{figure*}[htb]
 \centering
 \includegraphics[width=0.9\textwidth]{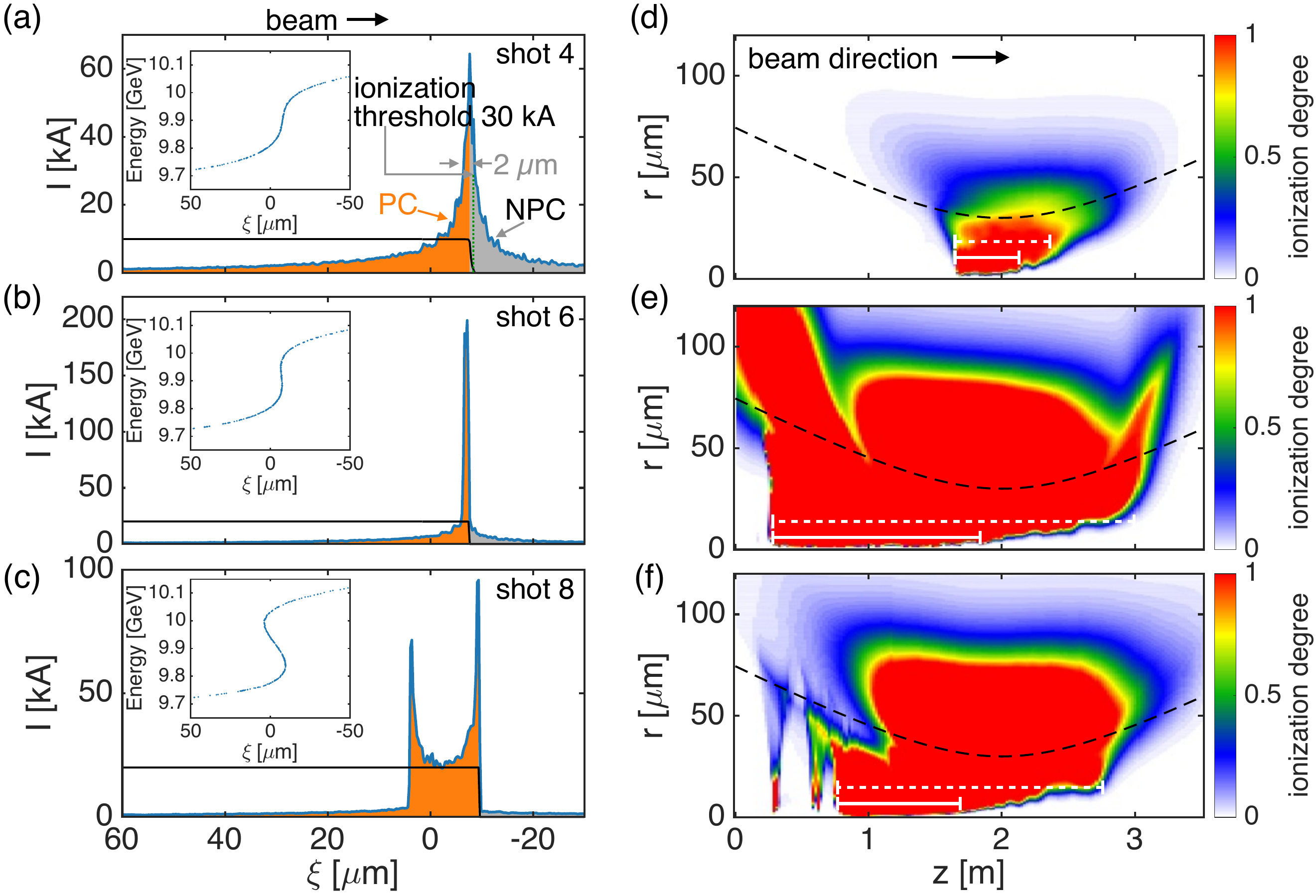}
 \caption{\label{fig1-plasma}Meter-scale plasma generation. (a) An example of a current profile derived from a slightly under compressed bunch [shot 4 from Fig. \ref{fig1-extreme-beam}]. The ionization degree, depicted by the black line, is calculated using the ADK model, showing ionization initiation at 30 kA that reaches full ionization within approximately 4 fs ($<$2 $\rm\mu m$). The grey shaded region indicates the nonparticipating charge (NPC), while the orange shaded region represents the participating charge (PC) following the ionization front. The inset displays the corresponding longitudinal phase space of the bunch. (b)-(c) Additional examples of a fully compressed and somewhat over compressed bunches arising from the RF jitter. (d)-(f) Spatial distribution of ionization degree of hydrogen molecule attained using the PIC code QPAD at 2 Torr gas pressure using the bunches in (a)-(c), respectively. The white lines indicate the length of the plasma (see text). The dashed black curves depict the spot size evolution in vacuum.}
\end{figure*}

The effects of variations in the magnitude and location of the current spike on the formation of beam-ionized hydrogen plasmas were investigated using QPAD. The results are shown in Fig. \ref{fig1-plasma}. Figure \ref{fig1-plasma}(a) shows the current profile of the drive bunch, with RF amplitude and phase settings similar to those used in the experiment [shot 4 in Fig. \ref{fig1-extreme-beam}]. The current profile often exhibits one or more (not resolved) femtosecond (fs) spikes, with an average peak current surpassing 60 kA and a mean length of $\leq$2 $\rm\mu m$ (FWHM). Below the prominent current spike, there exists a relatively lower current yet considerably broader charge distribution spanning over up to 100 $\rm\mu m$ with a peak magnitude of $<10$ kA.

The black solid line in Fig. \ref{fig1-plasma}(a) shows the ionization degree of the first electron of hydrogen molecules (IP=15.4 eV), plotted as a function of longitudinal position. This calculation was performed at the radial position where the transverse electric field of the beam attains maximum strength, utilizing a spot size of $\sigma_r=30~\rm\mu m$ and the ADK model \cite{ammosov_tunnel_1987}. This calculation underscores that the narrow current spike is sufficiently intense to rapidly ionize hydrogen molecules, leading to plasma production at the location of the spike. Ahead of the spike, the beam current falls below the threshold (30 kA) required for hydrogen molecule ionization. Consequently, the energy contained in this part of the bunch is wasted and appears at the initial energy point (nominally 10 GeV) on the spectrometer. We call this charge and any additional charge that happens to be at the location in the beam where the longitudinal electric field of the beam-induced wake is zero and therefore does not lose energy as the non-participating charge [NPC, marked by the grey shaded region in Fig. \ref{fig1-plasma}(a)-(c)]. For this current profile, the percentage of NPC amounts to approximately 30\% of the total charge, consistent with the observations from the experimental data as we shall later see.

Also shown in Fig. \ref{fig1-plasma}(a-c) are the longitudinal phase space plots of the drive bunch for the nominal bunch compressor magnet settings that results in pulses that are under compressed as in Fig. \ref{fig1-plasma}(a) to optimally compressed, Fig. \ref{fig1-plasma}(b), to over compressed- Fig. \ref{fig1-plasma}(c). This case of over-compression leads to a current profile that has two spikes separated by about 15 $\rm\mu m$.

A set of 50 PIC simulations were conducted using the simulated current profiles [the first 50 shots in Fig. \ref{fig1-extreme-beam}(a)] to investigate plasma and wake formation in 2.0 Torr (one of the pressures used in the experiment) static-fill hydrogen gas. The PIC simulations self-consistently include the ionization of the hydrogen gas using the ADK model \cite{ammosov_tunnel_1987} and track the evolution of the drive bunch within the meter-scale self-ionized plasma. Molecular hydrogen has an ionization potential of 15.4 eV, larger than the 13.6 eV of hydrogen atoms. Therefore in these simulations, we employed atomic hydrogen but with an ionization potential of 15.4 eV to accurately mimic the ionization pathway of molecular hydrogen by an ultrashort beam \cite{zhang_ionization_2021,nie_cross-polarized_2022}. In the simulation, we utilized a moving window with dimensions of $z=10~c\omega_p^{-1}$ (beam direction) and $r=7~c\omega_p^{-1}$ (transverse direction), divided into 1000 and 400 cells along each direction, respectively, where $c\omega_p^{-1}$ represents the plasma skin depth for the normalized density of $n_e=6.48\times10^{16}~\rm{cm^{-3}}$. The current profile of the drive bunch, acquired from the beamline simulation, was imported into the QPAD simulation using a piecewise-linear approximation with a discretization step of $\Delta z=0.2~\rm\mu m$, ensuring the resolution of fine structures in the current profile. For simplicity and stability, the drive bunch was approximated by a cylindrically symmetric Gaussian distribution in the transverse plane, with a normalized emittance of 20 $\rm\mu m$ and focused to a transverse spot size of 30 $\rm\mu m$, matching the experimentally measured size. This approximation removes the slice misalignment presented in the beam distribution obtained from the beamline simulation due to coherent synchrotron radiation (CSR) in the final bunch compressor. 

Figure \ref{fig1-plasma}(d) displays the spatial distribution of gas ionization degree for the nominal current profile shown in Fig. \ref{fig1-plasma}(a). This result was obtained by stitching the radial ion density profile at a given longitudinal location (e.g., average of the slices between 30 to 50 $\rm\mu m$ after the current spike) within the moving simulation window for every time step. The black dashed line traces the evolution of the beam’s spot size in vacuum, revealing a 30 $\rm\mu m$ focus at $z=2$ meters. In this instance, hydrogen molecules were ionized over a span of 50 cm as indicated by the white solid line, which covers the region of plasma that is fully ionized and the on-axis un-ionized region is narrow ($<5~\rm\mu m$). Note that other criteria, for instance, the region where the transverse size of the fully ionized region is broader than the unionized channel, results in a longer length, as indicated by the white dashed line. Figure \ref{fig1-plasma}(e)-(f) exhibit the outcomes corresponding to the other two representative current profiles shown in Fig. \ref{fig1-plasma}(b)-(c). In all cases there is a $\rm\mu m$-scale on-axis region where no ionization occurs because the radial electric field is below the ionization threshold.

Figure \ref{fig1-plasma}(d)-(f) show that the drive bunch with peak current exceeding 60 kA can generate meter-scale plasmas. In experiments using tightly focused drive bunches, the wake excitation may be limited by beam head erosion. If the beam head erosion limited plasma length is less than the pump depletion length, then the total energy transfer from the drive bunch to the wake can be affected \cite{blumenfeld_energy_2007}. However, in simulations and experiments reported here, head erosion is not a concern due to the relatively gentle beam focusing optics ($\beta^*\geq50$ cm), much higher peak current and 10 GeV energy of the beam. In this scheme, the product of the decelerating gradient and plasma length (see Table \ref{parameters}) is sufficient to drive the beam electrons towards pump depletion.

\begin{table}
\caption{\label{parameters}Beam, plasma, and wake parameters derived from PIC simulations.}
\begin{tabular*}{\textwidth}{@{}l*{15}{@{\extracolsep{0pt plus
12pt}}l}}
\br
Shot number & $Q_{\rm NPC}$ fraction & Peak plasma density & Plasma length & Max $E_{\rm DEC}$ \\
  &   & [$10^{16}~\rm cm^{-3}$] & [m] & [GeV/m] \\
\mr
4& $30\%$ & 6.48 & 0.50$^a$ $\sim$ 0.75$^b$ & 14.0\\
6& $21\%$ & 6.48 & 1.57 $\sim$ 2.77 & 16.4\\
8& $9\%$ & 6.48 & 0.93 $\sim$ 2.03 & 28.1 \\
\br
\end{tabular*}\\
$^{a}$Length of the fully ionized region where the on-axis un-ionized channel is narrow ($<5~\rm\mu m$), indicated by the white solid lines in Fig. \ref{fig1-plasma}(d)-(f).\\
$^{b}$Length of the fully ionized region where the transverse size of the fully ionized region is broader than the un-ionized channel, marked by the white dashed lines in Fig. \ref{fig1-plasma}(d)-(f).\\
\end{table}

The key quantities relevant to the topic of this paper can be derived from these simulations and are summarized in Table \ref{parameters}, highlighting substantial variations arising from current profile fluctuations. For instance, nonparticipating charge ranges from 9\% to 30\% for shots where ionization takes place (NPC is 100\% if peak current of the spike falls below the ionization threshold). Notably, the minimum NPC fraction observed in the experiment was about 25\%. In beamline simulations, the initial current profile of the electron bunch produced from the photocathode likely differs from the actual profile in the experiment. With a peak current exceeding 60 kA, the electron bunch is able to singly ionize hydrogen molecules, albeit over varying distances from 0.5 to 1.6 meters. In contrast, electron bunches with peak currents below 60 kA or gas pressures lower than 0.08 Torr (see Fig. \ref{fig2-QPAD}) partially ionize the gas over few tens centimeters, yielding a much shorter plasma with nonuniform reduced densities. Consequently, the maximum decelerating field experienced by the drive bunch fluctuates from 6 to 28 GV/m. Despite the substantial shot-to-shot fluctuations in plasma generation and wake excitation observed in both simulations and experiments, we will demonstrate that crucial information, including pump depletion and energy transfer efficiency from beam to wake, can still be extracted from the experimental data and PIC simulations.

\subsection{Wake Excitation Modeled using QPAD Simulation}
\begin{figure*}[htb]
 \centering
 \includegraphics[width=1\textwidth]{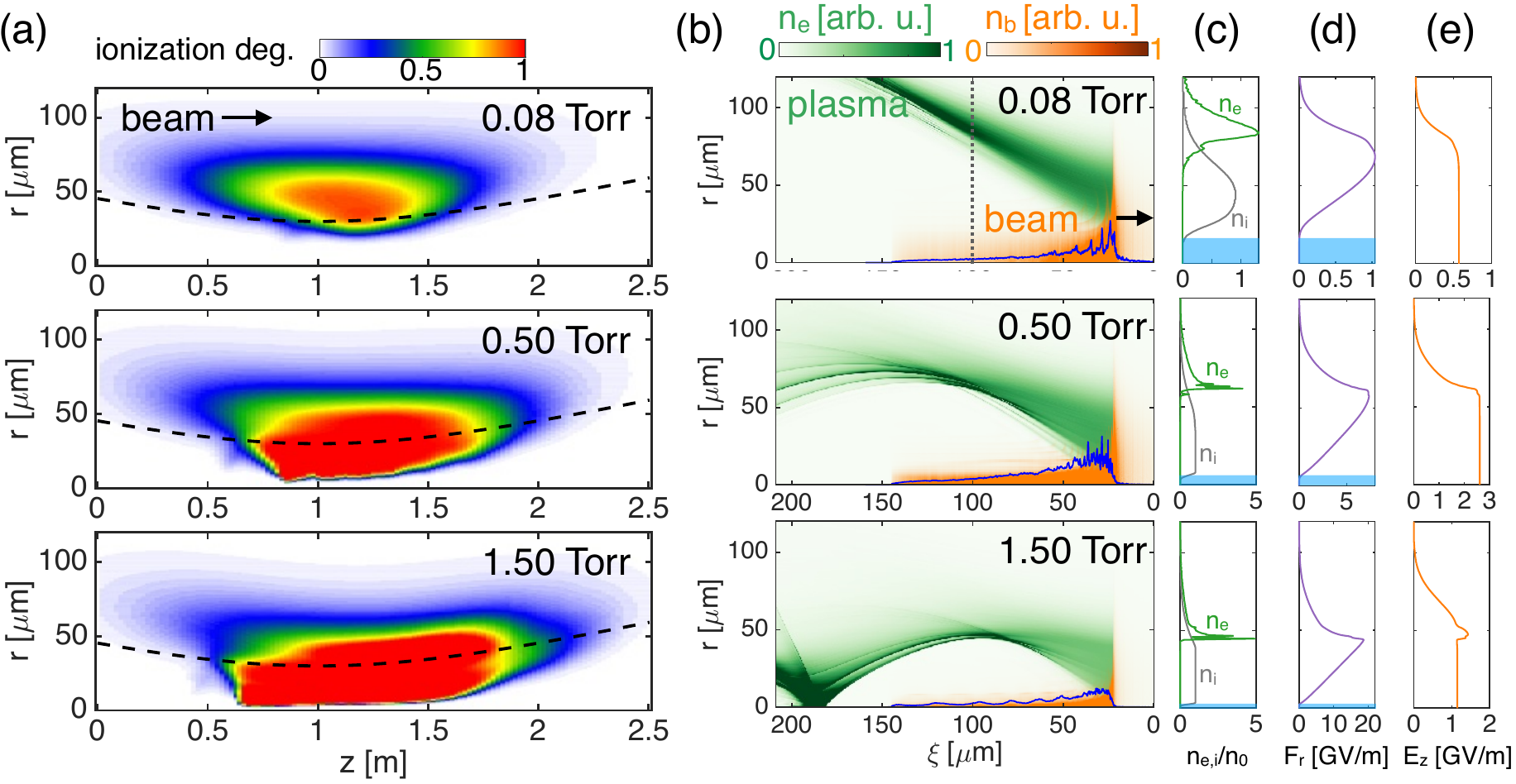}
 \caption{\label{fig2-QPAD}(a) Spatial distribution of ionization degree of hydrogen molecule obtained from QPAD simulation using the nominal beam with three different gas pressures. Experimental datasets were acquired at these and other pressures. The dashed black line shows the evolution of the beam size in vacuum. (b) Wakes excited by the drive bunch at the vacuum focus location. The blue lines show the on-axis density profile of the drive bunch. The peaks are due to betatron oscillations of these electrons. (c) The radial density profiles (green for electron and grey for ion, both normalized to the corresponding neutral gas density $n_0$) evaluated at $\xi=100~\rm\mu m$ [marked by the dashed line in the top row of (b)]. (d) and (e) show the transverse focusing field $F_r=E_r-B_\theta$ and the longitudinal field $E_z$ at the same slice. The shaded regions (blue) indicate the micro channel where the hydrogen gas is not ionized by the beam and contains a portion of the tail of the beam.}
\end{figure*}

Following the ionization of hydrogen gas, these beams can excite a long thin column of wake that is self-aligned to the plasma. Figure \ref{fig2-QPAD} illustrates various representative simulation outcomes employing the nominal current profile depicted in Fig. \ref{fig1-plasma}(a) at different gas pressures. Note that these simulations begin one meter upstream of the drive bunch's focal plane, as opposed to the prior two meters. This is reasonable since plasma with a density $>1\%$ of the fully ionized density is contained within this region. In Fig. \ref{fig2-QPAD}(a), we show the spatial distribution of the ionization fraction as a function of hydrogen pressure. Once the plasma formation is triggered by the large current spike, the plasma electrons begin to be expelled transversely and the plasma ions exert a focusing force on the subsequent beam slices following the ionization front. At the lowest pressure, the focusing force remains relatively mild, and the transverse spot size of the beam diminishes to roughly half its size in vacuum (illustrated by the dashed black line). With increasing plasma density, however, the focusing force exerted by the plasma ions intensifies, yielding a progressively narrower on-axis region that remains un-ionized.

Figure \ref{fig2-QPAD}(b) illustrates the density distribution of the drive bunch (orange) and the plasma (green) at these various densities. The hydrogen gas is rapidly ionized by the current spike, leaving the beam electrons in front of the spike unaffected. At the lowest density, the drive bunch primarily occupies the front of the wake, where the ionized electrons are blown out into the neutral gas region. The beam electrons experience only the rising but small decelerating field [0.5 GeV/m at the dotted line in Fig. \ref{fig2-QPAD} (b)], leading to a small energy loss to the wake. With increasing plasma density, the wake's length shortens, while the field strength intensifies, causing the drive bunch to lose energy at a higher rate. In the 0.5 Torr case, the back of the drive bunch extends to the center of the bubble where the longitudinal electric field changes sign. This represents a scenario where all the electrons among the participating charge suffer a varying degree of energy loss. At higher densities the wake wavelength becomes shorter relative to the drive bunch length. In all of these cases, two longitudinal slices of the beam can suffer the same energy loss because the decelerating field experienced by the beam is double-valued. Consequently, there can be a charge peak around the minimum energy (maximum energy loss) of the dispersed electron spectrum in the actual experiment. At still higher plasma densities (e.g., $>1.5$ Torr case), the charge comprising the back of the drive bunch crosses the point where the electric field of the wake changes sign (where the the bubble radius is the largest at roughly 100 $\mu m$) and extends into the accelerating phase, extracting energy from the wake which corresponds to energy gain on the energy spectrometer. These characteristic features of the wake are indeed reflected in the measured electron spectra as we shall later see.

The rest of Fig. \ref{fig2-QPAD} displays the radial density profiles of electrons ($n_e$) and ions ($n_i$), Fig. \ref{fig2-QPAD}(c), along with the radial variations of the transverse focusing Fig. \ref{fig2-QPAD}(d), and longitudinal decelerating field Fig. \ref{fig2-QPAD}(e) for the drive bunch evaluated at $\xi=100~\rm\mu m$. As plasma density increases, the focusing field of the wake intensifies, causing the low-density tail of the drive bunch to contract in the transverse direction. Consequently, the un-ionized region (marked by the blue shaded area) becomes narrower. The proportion of drive bunch charge within these channels diminishes from 22\% for the 0.08 Torr case to 2.5\% for the 1.5 Torr case. Therefore,  this small fraction of the electrons enclosed in the un-ionized channel region has negligible impact on the beam-to-wake energy transfer efficiency. Importantly, as shown in Fig. \ref{fig2-QPAD}(d) and (e), the focusing field is zero within the unionized channel (see the regions marked in light blue) and the decelerating field is uniform and has the same value inside as that outside the channel, making it suitable for positron acceleration \cite{gessner_demonstration_2016}.

\subsection{Experimental setup}
\begin{figure*}[htb]
 \centering
 \includegraphics[width=1\textwidth]{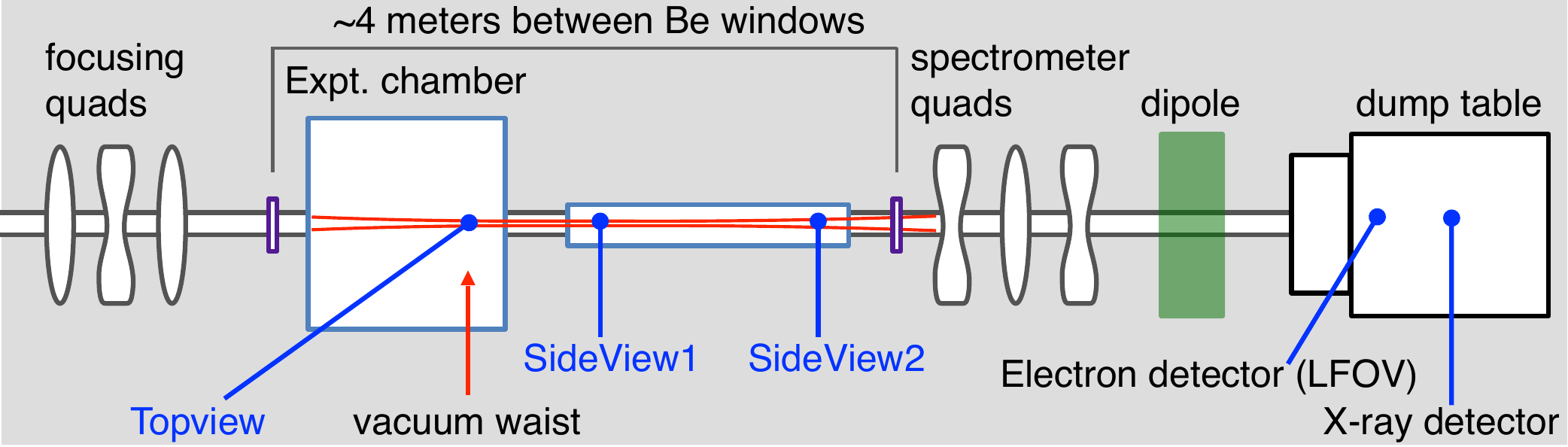}
 \caption{\label{fig3-expt-setup}Sketch of the FACET-II experimental area and relevant diagnostics for the results presented in this paper. The 10 GeV electron beam propagates from left to right and undergoes focusing by two (only one set is shown) sets of focusing quadrupoles into the experimental chamber. The region between the two beryllium windows with self-drilled holes is filled with hydrogen gas at an adjustable pressure ($<5$ Torr). Following interaction with the beam (field)-ionized plasma, the electron beam is dispersed and imaged  by an energy spectrometer, comprising finely tunable imaging quadrupoles, a coarsely adjustable dispersive dipole, and a large field of view (LFOV) phosphor detector. An x-ray detector measures the integrated betatron x-ray signals. Three cameras, Topview, Sideview1 and 2 collect the time-integrated plasma light in the visible portion of the electromagnetic spectrum.}
\end{figure*}

The FACET-II experimental area is described in detail in \cite{storey_wakeeld_2023}. Here in Fig. \ref{fig3-expt-setup}, we present a sketch of the setup and focus on the relevant diagnostics for the results presented in this paper.

The compressed 10 GeV electron beam was focused by two sets of quadrupoles into a static fill of hydrogen gas with a tunable pressure between 0.07 Torr to 2.2 Torr (maintained within 1\% precision). The gas was confined by two 50 $\rm\mu m$ thick beryllium windows, with holes drilled {\it in situ} by the electron beam itself, spaced approximately four meters apart. The slow rate of flow of hydrogen gas out of these small holes in the Be windows is matched by a continuous flow of gas into the system to maintain a constant gas pressure between the Be windows. The rate of flow through this system is not expected to product a significant density gradient from the upstream to downstream beryllium window. The gas flowing out of the Be windows is pumped out by a four-stage differential pumping system that is used to reduce the beamline pressure back to ultra-high-vacuum ($10^{-9}$ Torr) in the accelerator from the four-meter-long hydrogen gas region between the two beryllium windows which formed the hydrogen gas chamber. The beam was focused at the location monitored by the topview camera (downstream side of the experimental chamber) to a round spot of $\sim$30 $\rm\mu m$ with $\beta^*\approx50$ cm.

As the transverse electric field of the electron beam surpasses the hydrogen molecule field ionization threshold, a plasma is rapidly formed. The resulting self-emission from this plasma was observed by three cameras covering a total distance of approximately 1.5 meters. Plasma light was evident on all three cameras for hydrogen gas pressures exceeding 0.07 Torr, which indicates the formation of meter-scale plasmas. However, this does not necessarily imply that the plasma density, temperature or duration of the plasma emission was the same at these three locations. After passing through the plasma, the beam was directed to downstream diagnostics, enabling measurements of parameters such as charge, energy spectrum, and emittance \cite{green_beam_2018}. The charge entering the hydrogen chamber was measured by a toroid positioned upstream of the chamber. All these diagnostics operated on a single-shot basis \cite{gessner_facet-ii_2021}.

\section{Experimental Results}
\subsection{Energy Loss as a Function of Gas Density}
In the experiment, the pressure of the static-fill hydrogen gas was varied between 0.07 and 2.2 Torr- the upper range of pressure being limited by the differential pumping system. As the electron beam interacted with the plasma, its longitudinal slices underwent energy loss to or gain from the wake. The energy spectrum of the electron beam after interaction was recorded by the large field of view (LFOV) screen of the imaging spectrometer.  For each hydrogen pressure setting, we recorded approximately 200 consecutive shots to ensure reliable data collection.

\begin{figure*}[htb]
 \centering
 \includegraphics[width=1\textwidth]{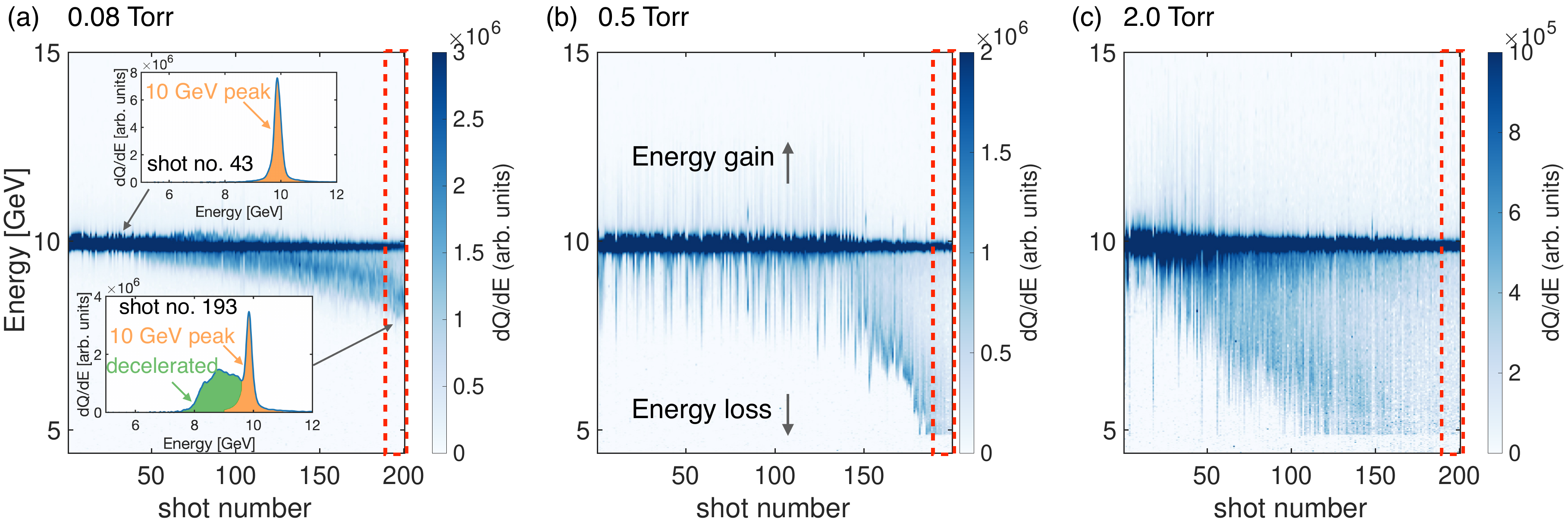}
 \caption{\label{fig4-spectrum}Linearized energy spectra of three representative datasets obtained at varied pressures. Each column represents a single-shot measurement. Each dataset was sorted by the total energy loss (in increasing order) of the bunch. Insets in (a) show two illustrative spectra, highlighting the 10 GeV peak (orange) and the decelerated segment of the beam (green). The shots enclosed by the dashed red rectangles in (a)-(c) are used to calculate the energy deposited in the plasma. For the 0.08 Torr shots the total charge with and without beam propagation through plasma is conserved, whereas for (b) and (c) a significant number of electrons with $<$5 GeV energy are not captured by the spectrometer.}
\end{figure*}

Figure \ref{fig4-spectrum} presents ``waterfall plots'' of 200 linearized energy spectra for the drive bunch across three datasets obtained at varying gas pressures. The camera that monitors the LFOV screen was covered by an ND2 filter so as to avoid possible  saturation of the detector by the large amount of charge in the 10-GeV peak to allow for full charge reconstruction. Each column within a dataset represents a single shot measurement (integrated over the undispersed direction). The 10 GeV peaks observed across all datasets suggest the presence of electrons that did not interact with the plasma. The signal below the 10 GeV peak corresponds to drive bunch energy loss. Unsurprisingly, all three datasets exhibit considerable fluctuations in energy loss, consistent with substantial variability in drive bunch parameters- most likely the peak current of the spike and its location within the bunch, as well as bunch length of the low current bunch structure, crucial factors in hydrogen gas ionization and energy loss to the wake. Despite these fluctuations, a clear trend of larger energy loss (electrons with energy $<10$ GeV) with increasing pressure is evident, indicating that a larger wake amplitude-length product resulted as the hydrogen gas pressure was increased. It is worth noting that the shot-to-shot charge variation in these datasets was about 2\%, namely, charge variation cannot account for the energy loss variation on a shot-to-shot basis. Also note that at and beyond a pressure of 0.5 Torr, there is some energy gain signal (electron signal with energies $>$10 GeV).

During these data runs, the imaging energy of the spectrometer was set to 5 GeV but the dipole magnet was set such that electrons with energy less than 5 GeV were not captured by the spectrometer, however, the measurement of lower charge is possible with this diagnostic down to $<$2 GeV through the use of reduced dipole strength as we shall see later. We segment the energy spectrum recorded by the spectrometer into two distinct portions: the 10 GeV peak and the decelerated part (5-10 GeV), as exemplified in the insets of Fig. \ref{fig4-spectrum}(a). In the case of shot no. 43, a single peak at 10 GeV is evident, indicating that the drive bunch experienced no energy loss, likely due to the peak current of the beam not reaching the required threshold for ionizing hydrogen molecule. Conversely, for another example (shot no. 193), the energy spectrum exhibits a large but reduced magnitude peak at 10 GeV (orange) alongside a broader peak at lower energies (green) which represents the decelerated segment of the drive bunch. In both cases, the 10 GeV peak corresponds to the non-participating charge, which can be distinguished and separated from the decelerated segment by applying a Lorentzian fit to the spectrum. The minimum NPC fraction ($Q_{\rm NPC}/Q_{\rm total}$) observed in the experiment was $\sim25\%$.

\subsection{Direct and Deduced Evidence for Energy Depletion of the Drive Bunch}
To attain the highest energy transfer efficiency from the drive bunch to the wake, a significant depletion of the drive bunch energy is necessary. The most straightforward way to observe energy depletion is to demonstrate that the minimum energy recorded by the spectrometer approaches zero. For instance, a recent work shows that a fraction of a 500-MeV electron bunch loses almost all its energy by driving wakes in discharge plasmas in a 20-cm-long capillary\cite{pena_energy_2023}. However, the beamline after the interaction point can effectively focus and transport electrons only within a certain energy range. For instance, at FACET-II, electrons with energy $<$1 GeV in particular, are prone to loss during the transport due to their substantial divergence.

\begin{figure*}[htb]
 \centering
 \includegraphics[width=1\textwidth]{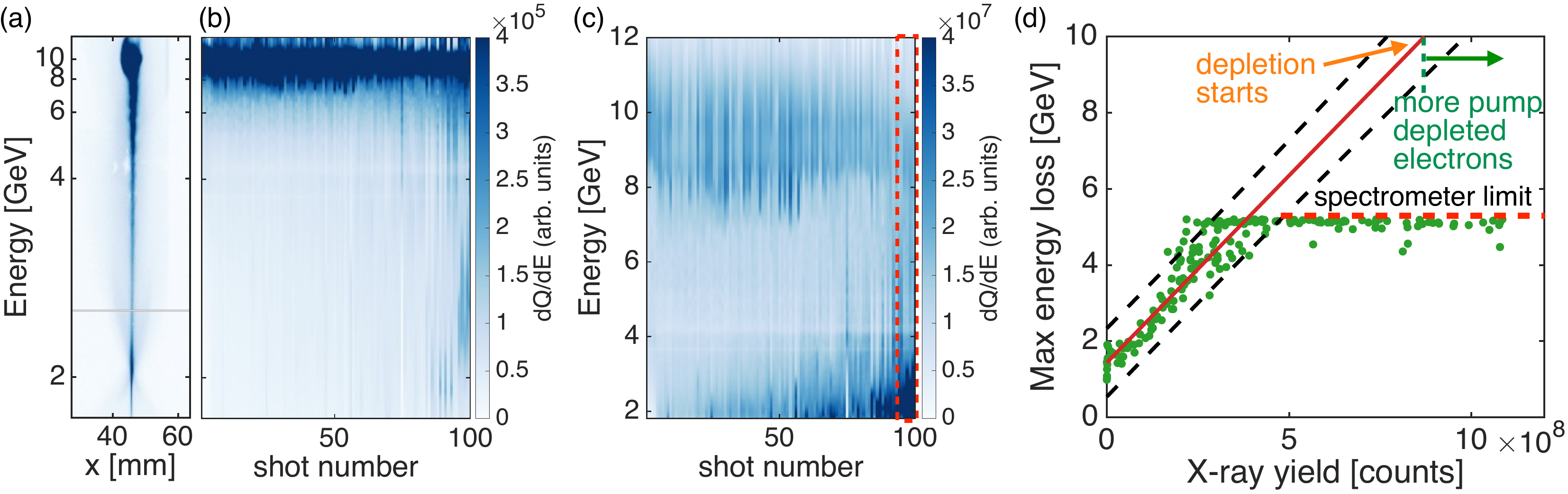}
 \caption{\label{fig5-pump-depletion}Direct (a)-(c) and deduced (d) evidence of some electrons approaching pump depletion. (a) A representative energy spectrum (raw data before linearization of the energy axis) recorded at a pressure of 2.17 Torr but with a lower dispersion setting of the spectrometer dipole shows that the energy of some of the decelerated charge indeed extends to below 2 GeV. Note that the 10 GeV peak is saturated on the LFOV screen. (b) A dataset of 100 consecutive shots at 2.17 Torr. (c) Linearized spectra of the dataset shown in (b), sorted by increasing energy loss of the electrons. The NPC appears as a broad peak after linearization due to saturation of the signal. Also note that the quads were set to focus at 2 GeV, broadening the 10 GeV peak further. The shots enclosed by the red dotted rectangle are used to calculate the energy deposited in the plasma and for estimating the energy transfer efficiency. Note that 73 out of 100 of the shots shown in (b) and (c) show decelerated electrons with energy extending below 2 GeV. (d) Maximum energy loss observed on the spectrometer plotted against betatron emission-x-ray yield obtained in a narrow cone angle in the forward direction (Fig. \ref{fig3-expt-setup}) for a different dataset taken at 2.0 Torr [Shown in Fig. \ref{fig4-spectrum}(c)]. For this dataset the spectrometer only captured electrons with an energy loss smaller than 5 GeV (indicated by the dashed red line). The red line represents a linear fit to the correlated data for x-ray yield less than $5\times10^8$ counts and energy loss less than 5 GeV. If this correlation continued up to 10 GeV energy loss, the intersection of the linear fit with the 10 GeV line would indicate the onset of pump depletion. The dashed black lines indicate the 95\% confidence interval of the fit. A larger x-ray yield beyond the intersection of the blue dotted line and the x-ray counts implies more energy-depleted electrons.}
\end{figure*}

In Fig. \ref{fig5-pump-depletion}(a) we show an example of the measured energy spectrum at a pressure of 2.17 Torr but with a less dispersive spectrometer dipole setting which allowed us to record on the LFOV sub-5 GeV electrons. The raw dataset of 100 shots before linearization of the energy axis is shown in Fig. \ref{fig5-pump-depletion}(b). These data show that the spectrum indeed extends to below 2 GeV, providing direct evidence of approaching pump depletion at this pressure. The linearized spectra for the whole dataset are shown in Fig. \ref{fig5-pump-depletion}(c). Note that after linearization, the low-energy signal becomes more prominent. In 73 out of the 100 shots, the charge of electrons with $E<3$ GeV exceeds 100 pC, indicating that for a plasma density of $7\times10^{16}~\rm cm^{-3}$ these 73 shots approached drive bunch pump depletion.

Now we provide supporting evidence of energy depletion in our experiment from the x-ray diagnostic \cite{p_san_miguel_claveria_commissioning_2023}. In Fig. \ref{fig5-pump-depletion}(d), we present the correlation between the maximum energy loss recorded by the spectrometer and the total x-ray yield in the forward direction mainly due to betatron motion of the electrons in the wake’s ion cavity. This data is from energy loss shown in Fig. \ref{fig4-spectrum}(c) where the maximum energy loss limited by the setting of the dipole dispersion is 5 GeV. Although the energy loss data is only available up to 5 GeV, the x-ray emission data is available for all shots. A linear correlation between the energy loss and x-ray emission is therefore only observed up to 5 GeV followed by a constant energy loss for higher values of x-ray signal. The red line in Fig. \ref{fig5-pump-depletion}(c) shows the best fit to the data points with an energy loss up to 5 GeV. Here we only show the 2.0 Torr dataset for clarity but all three datasets in Fig. \ref{fig4-spectrum} show a similar trend albeit with different slopes since the maximum decelerating field and plasma length are pressure dependent. The maximum energy loss corresponds to the maximum decelerating field experienced by one slice of the beam throughout the plasma length at a given plasma density. 
In contrast, apart from the wakefield strength (focusing force) and plasma length, the betatron x-ray yield also depends on variables such as beam energy and particle displacement from the beam propagation axis. Despite these complex inter-dependencies of x-ray yield with beam, wake, and plasma parameters, we hypothesize that the observed linear correlation of the betatron x-ray yield and the maximum energy loss up to 5 GeV can be extrapolated up to 10 GeV, as indicated by the red line representing the best fit to the data.

With this premise, the point where the extrapolated red line reaches the 10 GeV energy loss denotes the onset of pump depletion, i.e., signifies that some of the beam electrons have lost all 10 GeV energy while experiencing the largest decelerating field. Extending the same hypothesis, any further increase of the x-ray yield is interpreted as more electrons reaching pump depletion.

\section{Comparison between PIC simulations and Experiments}
\begin{figure}[htb]
 \centering
 \includegraphics[width=0.6\textwidth]{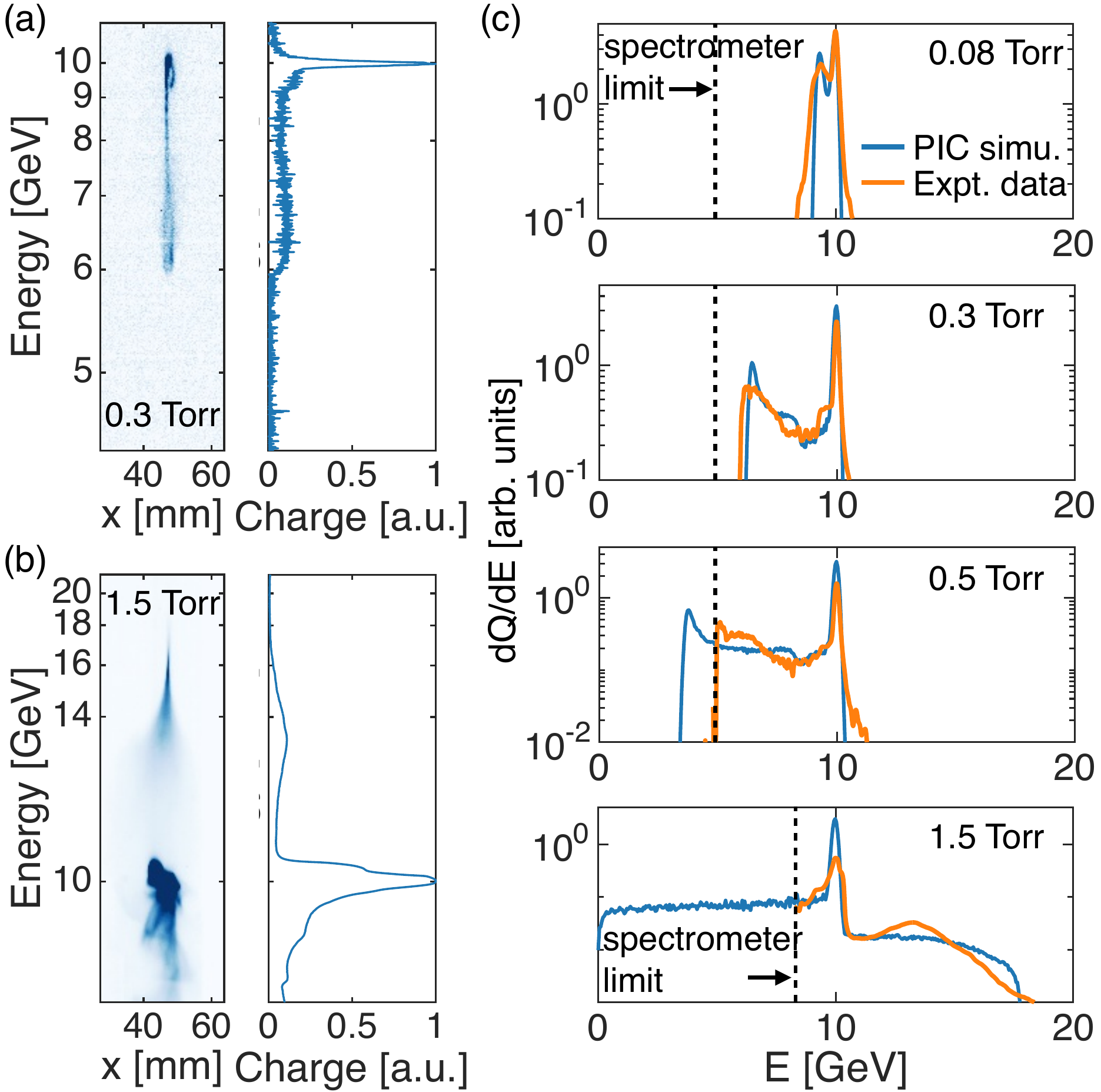}
 \caption{\label{fig6-simu-expt}Energy loss and gain of the drive bunch: comparison with PIC simulations. (a) Energy spectrum of an example shot from a 0.3 Torr dataset. Left panel: raw image from the spectrometer. Right panel: x-integrated energy spectrum. (b) Another example spectrum from a 1.5 Torr dataset with the spectrometer set to highlight the accelerated charge. Simultaneous measurements of energy loss portion of the spectrum were not taken. (c) Energy spectrum of the nominal drive bunch [see Fig. \ref{fig1-plasma}(a)] after interaction with hydrogen gas at varying pressures from PIC simulations (blue lines) and experiments (thick orange lines). The dashed black lines show the spectrometer limit (electrons with energy below this limit were not captured) for these datasets (note that the dipole setting for the 1.5 Torr case in the bottom row is different in order to measure electrons that have gained energy). The imaging energy was set to 5 GeV for the top three rows and 18 GeV for the bottom row in (c).}
\end{figure}

Up until now we have discussed the observed pressure-dependent energy loss of the drive bunch. However, as shown in Fig. \ref{fig2-QPAD}, when the pressure becomes $>0.5$ Torr some of the tail electrons begin experiencing the accelerating phase of the wake and therefore we expect to see the onset of energy gain in addition to energy loss on the spectrometer. Qualitatively, this is indeed what is observed in Fig. \ref{fig4-spectrum}(b), (c) and Fig. \ref{fig5-pump-depletion}. Since the beam current profile changes dramatically and is unknown in real-time data and we have used one particular current profile- that shown in Fig. \ref{fig1-plasma}(a)- to simulate the energy gain and loss results as a function of gas pressure any quantitative agreement is somewhat fortuitous. The results are presented in Fig. \ref{fig6-simu-expt}.

In Fig. \ref{fig6-simu-expt}(a), an example of raw energy spectrum of the decelerated bunch from the 0.3 Torr dataset is displayed. The spectrum indicates a minimum energy of 6 GeV (equivalent to a maximum energy loss of 4 GeV). At this low density, the nonlinear wake wavelength is relatively long such that the entire bunch experiences only deceleration. In contrast, Fig. \ref{fig6-simu-expt}(b) shows a spectrum from the 1.5 Torr dataset, where the spectrometer imaging energy is set at 18 GeV. The smallest size appears to be at 16 GeV because there is very little charge at 18 GeV [see the orange line in the bottom row of Fig. \ref{fig6-simu-expt}(c)]. Nevertheless, this result shows that some charge has indeed gained 8 GeV energy from the wake. Figure \ref{fig6-simu-expt}(c) shows the linearized energy spectra for four representative shots obtained at different gas pressures. The experimental data is illustrated by the orange lines. Across all cases, the prominent 10 GeV peak stems from the NPC. For datasets with pressures below 0.3 Torr, the lower-energy peak corresponds to the maximum decelerating field. As we have explained, this peak results from the double-valued decelerating field experienced by the drive bunch. The 5 GeV cut-off in the 0.5 Torr data arises from the spectrometer limit. In the 1.5 Torr case, the accelerated segment of the energy spectrum features a slight peak at 14 GeV and extends up to 18 GeV. Such a peak at 14 GeV could result from the drive bunch having a similar but smaller secondary charge peak at a particular longitudinal location. The simulated energy spectra, depicted by the blue lines, exhibit a reasonable agreement with the observed spectra. This comparison suggests that the longitudinal phase space distribution or the current profile we used for our QPAD modeling was likely the most frequent profile in the experiments.

\section{Energy Deposited into the Wake and Beam-to-Wake Energy Transfer Efficiency}
\subsection{Energy Deposited by the Beam into the Plasma Wake}
In PWFA, energy transfer from the drive to the trailing or accelerating beam occurs in two steps: First, the driver deposits its energy by exciting a large-amplitude wake and second, the particles in the trailing bunch extracts energy from this wake as it gains energy. In the blowout regime of PWFA reached in previous experiments that used a Li plasma \cite{litos_high-efficiency_2014,litos_9_2016,corde_multi-gigaelectronvolt_2015}, the (nonevolving) wake, once formed, does not change as the drive beam propagates through the plasma and eventually approaches pump depletion. This makes it rather straightforward to determine the energy deposited/extracted into/from the wake per unit length over the length of the wake. During this data run on FACET-II, we have used just  one (drive) bunch to estimate the energy transfer efficiency from the driver to the wake when at least a portion of the beam charge was fully depleted of its energy.

Utilizing the raw data such as that presented in Fig. \ref{fig4-spectrum}, we can calculate the energy deposited into the plasma by the drive bunch directly for shots where the maximum energy loss is $<$5 GeV (i.e., for hydrogen pressure $<$0.5 Torr). For these data the drive bunch propagation is neither beam-head erosion nor pump depletion limited. This is done by integrating the energy spectrum to quantify the energy carried by the beam recorded by the spectrometer at the end of the transport line. The charge of the original beam was measured by a toroid upstream of the gas region, allowing us to calculate the total initial energy contained in the incoming 10 GeV beam (16 J at 1.6 nC of total charge). We find that for a maximum energy loss up to 5 GeV, all electrons (within a measurement error of $<$2.5\%) reach the spectrometer after they are dispersed after interaction with the plasma. Assuming that these highly relativistic electrons only couple their energy to the wake, we evaluated the energy deposited in the plasma for ten shots enclosed in the dotted dashed rectangle in Fig. \ref{fig4-spectrum}(a), by subtracting the final energy measured on the spectrometer from the initial beam energy. We then calculated the percentage efficiency of energy transfer to the wake, from the fraction of the energy actually deposited in the plasma divided the maximum energy available to be deposited in the plasma times 100. Note that the efficiency calculation was done on a shot-to-shot basis since the NPC thus the available energy fluctuates. This procedure works well for energy loss data taken below hydrogen pressures of 0.5 Torr (see Fig. \ref{fig4-spectrum}).

\begin{figure}[htb]
 \centering
 \includegraphics[width=0.5\textwidth]{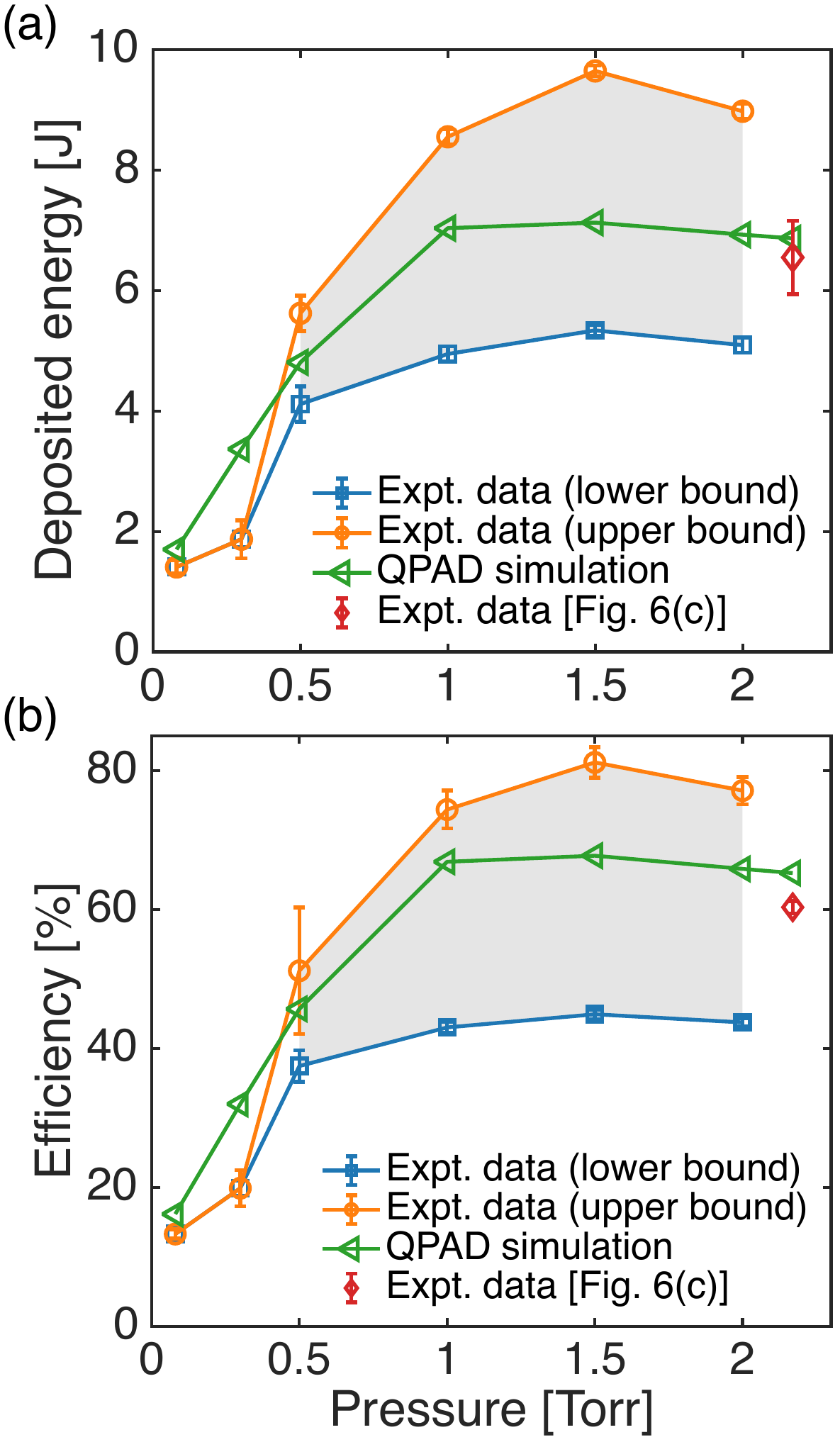}
 \caption{\label{fig7-efficiency}Deposited energy and effective beam-to-wake energy transfer efficiency vs. gas pressure. (a) Deposited energy. The blue and orange lines represent the lower and upper bound of deposited energy calculated using the spectrometer data. Each data point represents the average of 10 shots with largest deposited energy and the error bars indicate the standard deviation. For datasets with pressure $\leq0.3$ Torr, the two curves overlap since charge loss during beam transport after interaction did not happen. For higher pressures, the unknown energy of the missing charge introduces separation between the two estimates (lower and upper bound) where the actual deposited energy falls in the grey shaded region. The green line depicts the results obtained from QPAD simulations using the nominal current profile [see Fig. \ref{fig1-plasma}(a)]. (b) Beam-to-wake energy transfer efficiency. The red diamonds in (a) and (b) show the deposited energy and efficiency  retrieved from the 2.17 Torr dataset shown in Fig. \ref{fig5-pump-depletion}(b).}
\end{figure}

When the maximum energy loss exceeds 5 GeV we estimate the maximum and minimum energy deposited into the plasma (for hydrogen pressure of 0.5, 1, 1.5 and 2 Torr datasets) as follows. We assume that the energy contained in the x-rays and plasma emissions is small enough to be neglected. We then first estimate the energy deposited by the electrons that have lost up to 5GeV energy as before. The second part is the energy deposited by those electrons that lose more than 5 GeV energy and therefore do not appear on the spectrometer screen [“the missing charge” (MC)].  We can estimate the upper and lower bounds for the energy these electrons deposit. To do this, we attribute two extreme energies contained in the MC. For instance, if we assume that the energy of MC is $E_{\rm MC}=5$ GeV, then the additional deposited energy would be $Q_{\rm MC}\times5$ GeV. Conversely, if we assume that MC has been fully depleted of energy, i.e., $E_{\rm MC}=0$, then the additional energy deposited into the plasma by the missing charge is $Q_{\rm MC}\times10$ GeV. These two estimates added together give lower (underestimate, shown by the blue curve) and the upper (overestimate, shown by the orange curve) bounds for the energy deposited into the plasma wake, respectively. The results are shown by the blue and orange curves in Fig. \ref{fig7-efficiency}(a), where each data point represents the average of the 10 shots with maximum deposited energy for that dataset (enclosed by dashed red rectangles in Fig. \ref{fig4-spectrum}) and the error bar indicates the standard deviation. For pressures higher than 0.3 Torr, these two curves deviate due to the unknown energy of the missing charge. It is not a surprise that the relatively large energy range not covered by the spectrometer (0-5 GeV) introduces a large uncertainty in the deposited energy (the grey shaded region). The actual deposited energy should fall in the grey shaded region.

The green curve represents the QPAD simulation result. In these simulations, we used the nominal current profile depicted in Fig. \ref{fig1-plasma}(a) and changed gas pressure only. The simulation curve shows a similar trend with the experimental data and suggests a deposited energy of $\sim7$ J for datasets with pressure above 1.0 Torr.

The red diamond in Fig. \ref{fig7-efficiency}(a) represents the measured deposited energy retrieved using the 2.17 Torr dataset shown in Fig. \ref{fig5-pump-depletion}(b), where the energy spectra extend to below 2 GeV. For this dataset a lower estimate of the deposited energy was obtained by calculating the energy lost by the electrons with $E<$8 GeV range on the LFOV spectrometer since the 10 GeV peak in the raw images was saturated due to the removal of the ND2 filter so as to make the low-energy part of the spectra more prominent. The data point is the average of 5 shots with the largest energy loss and the error bar represents the standard deviation.

\subsection{Estimate of Beam-to-Wake Energy Transfer Efficiency at Pump Depletion}
Once we have the deposited energy (on a shot-by-shot basis), we can calculate the energy transfer efficiency from the beam to the wake by dividing the deposited energy by the total available energy (i.e., the energy carried by the NPC charge is omitted) for that shot and times 100. The results are plotted in Fig. \ref{fig7-efficiency}(b). Again, each data point with $\leq2$ Torr pressure corresponds to the average of 10 shots with maximum deposited energy for that dataset. The red diamond represents the efficiency retrieved from the 2.17 Torr dataset, where only the electrons with $E<8$ GeV were used in calculating deposited energy and efficiency. The green curve shows the QPAD simulation results using the current profile shown in Fig. \ref{fig1-plasma}(a). Note that when calculating the simulated efficiency, the energy carried by the 30\% non-participating charge has been omitted like we did in the experimental data analysis. The simulation results suggest that for datasets with a pressure above 1.0 Torr, the beam-to-wake energy transfer efficiency reaches about 65\%, which agrees reasonably with the 60\% efficiency retrieved from the 2.17 Torr dataset. These results therefore suggest that the beam-to-wake energy transfer efficiency has reached 60\% for pressures higher than 1.0 Torr (plasma density $>3.2\times10^{16}~\rm cm^{-3}$).

\section{Conclusion}
In conclusion, we have presented the analysis of the first experimental run by the PWFA Collaboration carried out at FACET-II. All experiments were conducted utilizing a single drive electron bunch. The linac setup produced time-structured beams that frequently featured one or more ultrashort, high-current spikes superimposed on a broader lower-current charge distribution. Our experimental observations and beamline simulations have revealed noteworthy fluctuations in plasma production and subsequent wake excitation, primarily attributable to variations in beam compression induced by RF jitter. Nevertheless, we have demonstrated the feasibility of utilizing such high current spike beams to field-ionize molecular hydrogen, thereby generating meter-scale plasmas. Furthermore, we showed the capability of generating large-amplitude wakes that, beyond a certain pressure threshold ($>$1.5 Torr), lead to energy depletion of a significant portion of the 10 GeV electrons while transferring approximately 60\% of the useful beam energy to the wake. We also provided evidence that certain electrons in the rear of the bunch can gain up to 8 GeV of energy from the wake. All experimental observations are reasonably replicated by PIC simulations conducted using the QPAD code.

\section*{Acknowlegement}
This work was supported at UCLA by the U.S. Department of Energy through Grand No. DE-SC0010064, and at SLAC by the U.S. Department of Energy under contract number DE-AC02-76SF00515. This research used resources of the National Energy Research Scientific Computing Center (NERSC), a U.S. Department of Energy Office of Science User Facility located at Lawrence Berkeley National Laboratory, operated under Contract No. DE-AC02-05CH11231 using NERSC award HEP-ERCAP-MP113. The authors thank Qianqian Su, Lance Hildebrand and Yujian Zhao for their help with the QPAD code development, as well as Lauren Alsberg, Max Gilljohann, Carsten Hast, Ryan Loney, Aime Matheron, and Marcellus Parker for their help with the experiment. A. Knetsch and P. San Miguel Claveria were supported by the France-Stanford Center for Interdisciplinary Studies for their travels to SLAC National Accelerator Laboratory. E. Gerstmayr was supported by the U.S. Department of Energy, Office of Science, Fusion Energy Sciences under Award DE-SC0020076.

\section*{References}

\providecommand{\newblock}{}

\end{document}